\documentclass{emulateapj}
\shorttitle{GRB AFTERGLOW APPARENT OPTICAL BRIGHTNESS DISTRIBUTION}
\shortauthors{Akerlof \& Swan}

\def\lsim{\mathrel{\rlap{\lower4pt\hbox{\hskip1pt$\sim$}}
    \raise1pt\hbox{$<$}}}                
\def\gsim{\mathrel{\rlap{\lower4pt\hbox{\hskip1pt$\sim$}}
    \raise1pt\hbox{$>$}}}                
\begin{document}
\title{AN ESTIMATION OF THE GAMMA-RAY BURST AFTERGLOW APPARENT OPTICAL BRIGHTNESS
  DISTRIBUTION FUNCTION}

\author{
Carl~W.~Akerlof\altaffilmark{1}~and
Heather~F.~Swan\altaffilmark{1}}

\altaffiltext{1}{University of Michigan, Randall Laboratory of Physics, 450
        Church St., Ann Arbor, MI, 48109-1040, cakerlof@umich.edu, hswan@umich.edu}

\begin{abstract}
By using recent publicly available observational data obtained in conjunction with the NASA \emph{Swift} gamma-ray burst mission and a novel data analysis technique, we have been able to make some rough estimates of the GRB afterglow apparent optical brightness distribution function. The results suggest that 71\% of all burst afterglows have optical magnitudes with $m_R < 22.1$ at 1000 seconds after the burst onset, the dimmest detected object in the data sample. There is a strong indication that the apparent optical magnitude distribution function peaks at $m_R \approx 19.5$. Such estimates may prove useful in guiding future plans to improve GRB counterpart observation programs. The employed numerical techniques might find application in a variety of other data analysis problems in which the intrinsic distributions must be inferred from a heterogeneous sample.
\end{abstract}

\section{Introduction}

One of the outstanding questions about gamma-ray burst afterglows is their
optical luminosity. Since the first counterpart \citep{c1} was identified in February 28, 1997, GRBs have been
detected optically over an intensity range that spans at least 14 magnitudes using
instruments ranging in aperture from 10 cm to 10 meters. Although the NASA \emph{Swift}
mission successfully determines celestial coordinates to accuracies that are often
better than a few arc-seconds, less than 50\% of all \emph{Swift} detections have led to
identifiable optical counterparts. The reason for this relatively low rate has been
the subject of much speculation. The three popular views are that: ({\it{i}}) GRBs are born in dusty, opaque
star-forming regions \citep{aaa1} \citep{aaa2}  \citep{aaa3} \citep{aaa4}
or,  ({\it{ii}}) originate at redshifts that make them invisible to us at optical wavelengths \citep{dark1} \citep{dark2} or
({\it{iii}}) are intrinisically dimmer than average \citep{bbb1} \citep{bbb2}.
No doubt, the truth is some combination of these possibilities.
We do not address these questions directly in this paper. Instead, we have tried
to estimate the fraction of bursts with afterglows that are reasonably accessible to detection with
observatories now in existence. We have developed a fairly simple procedure for
using the reported instrumental detection thresholds in conjunction with the actual
distribution of detected magnitudes to infer the underlying apparent afterglow optical
brightness function.

\section{ Overview of Analysis Technique}

There are now more than 200 \emph{Swift} GRB detections since the launch of this mission on November 20, 2004. The world community of ground-based astronomers has responded with optical observations of essentially all of these events, greatly augmenting the onboard measurements of the \emph{Swift} UVOT camera. From the data that have been reported, principally via the GCN, one can obtain the optical brightness for detected events, $m_{det}$, and the limiting magnitudes, $m_{lim}$, for those that are not. With this primary data, we have estimated the detected and limiting magnitudes at a fixed time of 1000 seconds post-burst and extrapolated the limiting magnitude data to include the detected events as well. For each of these steps, we will demonstrate that the statistical techniques appear to be quite robust. This is principally due to the fact that the analysis is based solely on cumulative probability distributions for $m_{det}$ and $m_{lim}$. Thus, estimation errors for individual events tend to get washed out in the mean as long as gross systematic effects are avoided. It is easy to see that one reason that less than 50\% of all bursts have detected optical counterparts is due to the limited sensitivities of the ensemble of instruments that was available at any given time. That can be framed more precisely by assuming that Nature has provided some intrinsic optical afterglow luminosity distribution to us on Earth, specified in magnitudes. For each GRB detected by \emph{Swift}, there is one best observational limiting magnitude. The convolution of these two distributions must be the observed distribution of $m_{det}$. This "equality" can be converted to an optimization problem of finding the best intrinsic afterglow distribution that satisfies this constraint. This estimate is probably the best we can do with the extremely heterogeneous observations that have been reported and the finite statistics of the sample.

\section{Data Selection and Correction}

The data for the ensuing analysis were collected from 118
\emph{Swift}-identified gamma-ray bursts that spanned a 447-day period from
February 15, 2005 through May 7, 2006. Both the GRB detection magnitudes and
limiting magnitudes were subjected to some identical selection criteria and
corrections. Foremost, the mid-point of the optical observations were required
to lie within a factor of 10 of a nominal post-burst time of 1000 s,
ie. between 100 and 10000 s. Observations were restricted to V or R band with
unfiltered counting as R. These restrictions eliminated 10 bursts from further
consideration; 9 due to the time cut and 1 due to the observing wavelength
($K_s$ band). Explicitly, a few events were labelled as non-detections when the only actual detections
evaded the allowed time window or filter constraints.
All magnitudes were also compensated for galactic absorption
using the NASA/IPAC Extragalactic Database Web-based
calculator\footnote{http://nedwww.ipac.caltech.edu/forms/calculator.html} that,
in turn, is based on the work of \citet{schlegel}.  For data taken under V
filters, a further adjustment of -0.41 magnitudes was applied to compensate for
the average GRB color difference between V and R. This adjustment is the
average difference between V and R for time periods ranging from 0.2 to 1 days
for 5 GRBs which had many measurements of V and R at many different times: 990510 \citep{vr1}, 021004 \citep{vr2}, 050502A \citep{vr3}, 020813 \citep{vr4} and 030329 \citep{vr5} \citep{vr6} \citep{vr7}.  The lightcurves were characterized by identical power-law decays so there is no evidence of chromatic variability over these time spans.

Beyond this point, the additional selection criteria for $m_{det}$ and
$m_{lim}$ somewhat diverged. For each of the 43 events with valid detections,
the measurement with an observation time logarithmically closest to 1000 s was
chosen. The list is displayed in Table~\ref{tab:detections}. (Much of the data for this paper was obtained from the GRBlog Web pages\footnote{http://grad40.as.utexas.edu/} maintained by \citet{c2} which enormously facilitated this project.) In order to proceed further, we must compare the optical brightnesses at a common post-burst time delay.

To make this project work, we needed to establish that it was possible to
extrapolate each observed magnitude at $t$ in the range [100, 10000] to a fixed
time, $t_c = 1000$ s. Fortunately, there was sufficient data for 37 of the 43
events to extract a power-law exponent, $\alpha$ for the temporal behavior of
each burst. With these values, we could make a reasonable estimate of $m_{det}$
at $t_c$. We also performed a similar calculation assuming a {\it fixed} value
for $\alpha = -0.70$. The two cumulative probability distributions for the
extrapolated values of $m_{det}$ are plotted in Figure 1. Application of the
Smirnov-Cram\'er-von Mises test shows that the two distributions are
effectively identical \citep{scvm1}, \citep{scvm2}. This gives us some confidence that the same power-law extrapolation is appropriate when the burst afterglows are NOT detected. This is verified by looking at the cumulative distributions of the observation times for the detections and non-detection upper limits (to be described below). This is shown in Figure 2. As expected, the detected events lie close to $t_c$ by virtue of the imposed selection criteria. The undetected events have no such bias. Nevertheless, their median lies close to 1000 s as well. We can make this more quantitative by comparing the RMS average magnitude shifts for the detections and upper limits due to translating from $t_{burst}$ to $t_c$. With $\alpha = -0.70$, the average detected magnitude
is shifted by 0.59 when extrapolating from $t_{burst}$ to $t_c$ while the similar number for upper limits
is 1.06.  Thus, the estimated cumulative distribution for the upper limits will be somewhat poorer but the plots in Figure 1 demonstrates that this is unlikely to be significant.

The estimation of the instrumental upper limits for afterglows, $m_{lim}$, is more complex.
First of all, very few research groups report $m_{lim}$ if there has been a detection.
Even if they do, there is a serious bias that will tend to shift $m_{lim}$ to greater values:
a large telescope is much more likely to observe a GRB if the optical counterpart has already
been announced. We have found a slightly devious way to get around these difficulties by using the
unbiased limiting magnitude distribution for non-detections to estimate the limiting magnitude distribution
for all bursts. For each undetected GRB, all limiting magnitude
reports are transformed as if they were detections to $t_c$, only requiring an observation time
within the [100, 10000] s window. The maximum magnitude of each set is adopted as $m_{lim}$ for
that burst. 65 events survived this analysis and are listed in Table~\ref{tab:undetections}.

Our task now is to create a distribution of all limiting magnitudes, both detected and undetected,
knowing only the values for the undetected. One obvious fact is that the limiting magnitudes
for detections will, on average, be deeper. In fact, if a detection is made at $m_{det}$, the
value for $m_{lim}$ will lie somewhere between $m_{det}+\sigma_{det}$ (where $\sigma_{det}$ is the
measurement error associated with $m_{det}$) and the best limiting magnitude ever
reported. If $m_{lim}$ truly represents the maximum sensitivities for the ensemble of bursts, the
simplest tactic is to take the median of the subset of $m_{lim}$ in the prescribed range and incorporate
that value into the entire set of $m_{lim}$. By performing this
recursively over the set of detected GRB afterglows, ordered by decreasing $m_{det}$, one can fill out the
otherwise missing entries.

We carried this one step further to better understand the stability of this method. We generated
1001 $m_{lim}$ distributions using a uniform random number generator to select the interpolated
values. For each successive element of $m_{det}$, a modified subset of $m_{lim}$ is considered
that includes all elements of $m_{lim}$ with values greater than $m_{det}+\sigma_{det}$ adjoined to
the lower limit value. A uniformly distributed random number then uses the cumulative distribution of the
restricted set to select an appropriate random value to be adjoined to $m_{lim}$.
In the limit of a large sample of $m_{lim}$ distributions, all possible sets for $m_{lim}$ will be generated
consistent
with the constraints imposed the values for the undetected $m_{lim}$ and the detected $m_{det}$.
To recover the best estimate for $m_{lim}$, the 1001 distributions were individually ordered by value. To select
the 108 elements of $m_{lim}$, the first value was chosen as the median of the set of first values of the
1001 Monte Carlo sets, the second value from the set of second values, etc. A similar procedure defines
the first and third quartile distributions.
If the distribution of such sets is tightly confined, we have reason to anticipate that this is
an adequate approximation of reality. The results are shown in Figure 3. The median distribution lies
within tight bounds constrained by the first and third quartiles.

The validity of this procedure was verified by modeling this deconvolution process assuming
knowledge of the true $m_{lim}$ distribution. For sake of computational simplicity, the $m_{lim}$
cumulative distribution was approximated by a Fermi-Dirac distribution with the two free parameters
chosen to best fit the apparent shape inferred from the analysis described above. The $m_{det}$
distribution was
taken from the 4-parameter b-spline representation described in Section 4 below. This allowed us to
create for $N$ events, a list of simulated Monte Carlo GRBs with values for the afterglow and
limiting
instrumental detector magnitudes determined by the two assumed cumulative distributions. Comparing
the two values, event by event, generated two
sub-samples: the `detected' events for which the afterglow was brighter than the instrumental
limit and the `undetected' events for which the opposite was true. The Monte Carlo samples
reproduced the detected/undetected event ratios essentially exactly. Applying the deconvolution
scheme that has been described, we found excellent agreement with the input assumptions for the
distribution of $m_{lim}$. One reason for the stability of this technique is the broad dispersion
of sensitivities of ground-based instruments reporting results. One measure is the distribution of
apertures: it is approximately logarithmic from 0.2 to 8.2 meters with
$dN \propto d({\rm aperture})/{\rm aperture}$.

Figure 4 shows the histogram distributions of detected GRBs and the limiting magnitudes for non-detections, both scaled to $t_c$ = 1000 s. The distributions are roughly similar with the latter
edging just a bit deeper. Such rough equality is what one might naively expect for the situation in
which about half of all events evade detection. Above $m_R = 19$, there are twice as many
non-detections (29) as detections (14).

\section{Finding the Optical Brightness Distribution Function}

The basic idea of this calculation is to specify the apparent optical brightness
function by a small set of parameters and, with this input, estimate the magnitude
distribution of detected events modulated by the actual probability of making such a
set of measurements with the required threshold sensitivity. By the usual least squares
techniques, the parameter set describing the afterglow brightness function is adjusted so that
the predicted distribution of detections closely matches the actual measurements. With
that in mind, we originally set out to represent the integral brightness distribution
function, $F(m)$, by a set of cubic b-splines uniformly spaced over the range of observed
magnitudes. Working with the integral distribution function removes the ambiguity
of selecting the binning interval that is implicitly required for defining the
associated differential distribution. However, the tradeoff is that the representation
of the integral distribution must guarantee that the function is monotonic over its
entire range. In detail, it was realized that computing $F$ as a function of the magnitude, $m$, led to problems near the endpoints where $F$ must approach either 0 or 1. Inverting the representation so that $m(F)$ is described by uniform b-splines over the interval, [0, 1], takes care of the endpoint problem nicely although at the expense of denying solution by linear regression.

Despite some misgivings about poor computational speed, it was found that the downhill simplex minimization method of \citet{c3} was quite capable of finding solutions quickly for spline curves defined by up to seven degrees of freedom. The IDL numerical analysis package\footnote{ITT Visual Information Solutions, ITT Industries, Inc. } was used for these computations, in particular the {\tt AMOEBA} routine adapted from section 10.4 of \emph{Numerical Recipes in C ($2^{nd}$ edition)}\citep{c5}. This approach made it convenient to enforce the monotonicity of the integral distribution function - whenever an evaluation of the goodness-of-fit function was requested with b-spline coefficients leading to zeros or negative values of the distribution function derivative, $dm/dF$, the returned value was set to exceed the maximum of all previous values over the simplex. Thus, non-monotonic integral distributions were easily rejected along with other computational problems.

As sketched above, we fold the estimated detection limiting magnitude distribution with a parametrically
defined function describing the true GRB afterglow distribution to predict the observed distribution of actual
detections. The starting point for this calculation is the integral distribution of detection upper limits, $m_{lim}$, described earlier. This is a staircase function with uniform vertical steps between irregular intervals, $\Delta m = m_{i} - m_{i-1}$, in which the
probability of observing with a given limiting sensitivity, $p_i \equiv p(m_{i-1} \to m_{i})$, is uniform. Within each of these intervals of magnitude, the expected number of detected GRB events will increase by an amount, $\Delta f^{calc}_i = \Delta F_i \cdot p_i$, where $\Delta F_i = F(m_{i}) - F(m_{i-1})$ is the associated change in the optical brightness distribution function over $\Delta m$.
The sequence of values for $F(m_i)$ are computed by inversion of the cubic spline representation,
$m(F)$.

Once the set of $\Delta f^{calc}_i$ is constructed, the cumulative probablity distribution for
the expected number of detected events can be obtained by summation:
$f^{calc}_i = \sum^i \Delta f^{calc}_j$. A trivial modification of this procedure allows one to
compute $f^{calc}$ for the sequence of ordered values of $m_{det}$ that characterize the actual
GRB detections. The experimentally observed cumulative distribution for these events, $f^{obs}_i$,
is just a sequence of rational fractions, $(1, 2, 3, \cdots , n_{det})/n_{total}$ where $n_{det}$
is the number of detected GRBs and $n_{total}$ is the number of all events considered, detected and
undetected alike.
The strategy to optimize the shape of $F(m)$ is now fairly simple: form the differences,
$\delta_i = (f^{calc}_i - f^{obs}_i)$ and minimize the sum of squares, $\sum^{n_{det}} \delta_i^2$. This last quantity defines the least-squares goodness-of-fit function that drives the downhill simplex routine mentioned previously. The montonicity of the cumulative distribution function helps ensure the
stability of the optimum fit.

\section{Results and Discussion}

The calculation described above was carried out with 4 to 7 degrees of freedom
for the cubic spline representation, corresponding to dividing the range of $F$, [0, 1], into one to four equal segments. The resulting fit is shown in Figure 5 along with the actual GRB detections. The fits are qualitatively excellent.

The corresponding integral distribution function for the apparent optical
brightness is shown in Figure 6. The curves all follow the same shape. The
range of validity of these curves extends at least to the 90th percentile of
the $m_{det}$, 20.5. At this point, the cumulative intrinsic afterglow
distribution accounts for 57\% of all \emph{Swift}-identified GRBs. The most extreme useful point corresponds to the
deepest detection at $m_{det} = 22.1$ where the intrinsic
distribution reaches 71\%. The remaining 29\% may constitute two populations:
GRBs inside optically dense regions or at redshifts beyond the Lyman-$\alpha$
cutoff. Since our statistical method relies on actual detections, the 29\% could easily
be somewhat lower and details of the high-magnitude afterglow distribution cannot be resolved.
Similar conclusions about the population of dim or dark GRBs have been reached by others from far different
arguments \citep{dark1} \citep{dark2}. Thus, the original question of why half or less of
all GRBs are optically identified has been resolved by the realization that roughly 25\% are lost
because they are dimmer than $m_{det} \simeq 22$ and the rest are missed because the available
instrumentation is inadequate. This partially answers one of the issues that led us to
this analysis, our observations of the afterglow of GRB 060116 \citep{swan1}.
Within 2000 seconds, the afterglow became dimmer than $m_R \sim 22$, making it an exceedingly
difficult target for further measurements. It is apparent that some but not all of the missing
optical counterparts are due to such dim but detectable objects.

While recognizing that differentiation amplifies errors, it is still useful to look at the differential GRB afterglow magnitude distribution determined directly from the cumulative distribution discussed above. As shown
in Figure 7, a peak appears at $m_{det} \approx 19.5$ which is only slightly displaced from the peak in the actual observed $m_{det}$ distribution. One might argue that statistical errors in evaluating $m_{lim}$ could shift this somewhat rightwards but unitarity puts limits on how much further the integral distribution can rise without changing slope. Thus, the overall behavior of the apparent GRB afterglow distribution is likely to follow closely the curves shown. Some caution should be exercised
about over-interpreting the physical significance of this peak. Since the BAT detector on
{\it Swift} operates in flux-limited mode, cut-offs at low brightness may simply be a reflection
of a proportional correlation to lower fluxes in $\gamma$-rays.

We have described a statistical analysis of GRB optical afterglows that has attempted to
obtain the brightness distribution for observers on Earth to better understand the population
of dimmer events and the criteria for improving such investigations. By including the
distortion effects of instrumental characteristics and by comparing at a time accessible to almost all
observers, our results are largely biased only by the trigger threshold of the BAT detector onboard
{\it Swift}. A rather different approach has been attempted by two groups during the past
two years \citep{gendre1}\citep{nardini1}\citep{nardini2}. Their aim is to find discriminants that
would identify sub-classes of GRB events by translating observed fluxes to the rest frame of the GRB.
In particular, Nardini, et al. have found that by using
those events with redshift information, they could project the optical flux in R-band back to the
GRB rest frame at a proper time of 12 hours. For a typical burst with $z \sim 2$, this corresponds
to an observation 1.5 days following the burst trigger, $\sim 100$ times greater than the value
of $t_c$ of 1000 seconds employed in our analysis. At this late epoch, they find that the majority
of events are clustered in luminosity with a standard deviation of 0.70 magnitudes. A low-luminosity
population is also identified as a minority constituent of an apparently bimodal distribution
and exhibits a factor of 15 lower flux. In their most recent paper, they include 25 {\it Swift} bursts of which 17 are referenced in this present paper. The high-flux fraction of Nardini events
has a mean observer frame brightness about 1 magnitude greater than our entire detection
sample while the low-flux cluster, with only 4 events, is statistically indistinguishable. Given
the different methods and goals of the Nardini analysis, no further comparison is likely to
be meaningful.

\section{Implications for Future GRB Observations}

Observations of GRBs are difficult and expensive primarily because of the reliance on large X-ray and
$\gamma$-ray detectors in space such as {\it Swift} and GLAST, each of which costs a good fraction of a billion
dollars. Recent history has shown that multi-wavelength observations considerably enhance the amount of information
about these elusive events. At the present time, we still do not have a definite theory of the energy transport
within a GRB jet - it could be baryonic, $e^\pm$ pairs or electromagnetic Poynting flux. Many hope that if GRBs
are better understood, they could help improve our understanding of the early star-formation period of our
Universe. In any case, research is bound to continue in this area for many years to come although launching of
new space missions dedicated to GRBs will likely be infrequent. The analysis in this paper suggests that a natural
threshold sensitivity for optical observations of {\it Swift}-detected bursts is $m_R \approx 20$. The data
gathered for this paper show that such levels are routinely achieved by 2-m telescopes. The cost of such instruments
is in the neighborhood of \$5 M, especially if purchased in multiple units. The total number of such units can be
gauged by the following simple argument: the sky is dark above any specific site for about $\frac {1}{3}$
of the day, a randomly detected GRB will be at an immediately accessible zenith angle about $\frac {1}{3}$ of the
time and the weather at a good site will be suitable with probabiity of $\frac {3}{5}$. The joint probability of
all three independent conditions is $\frac {1}{15}$, implying that optimal coverage is achieved with $\approx$ 15 instruments globally distributed around the Earth. The overall optical detection probability is modified to some extent by details of $\gamma$-ray detector pointing constraints. Clearly, a number of areas on Earth are already
well populated with research-grade telescopes, particularly Chile and southwestern United States. Many parts of
the world are not so well blessed. Some nations such as Thailand and Iran have recognized the scientific niche for observing optical transients and expect to install 2-m optical telescopes within a few years. That still leaves
a number of sites in Asia and elsewhere that could successfully enhance global coverage of rare phenomena such
as GRBs. An alternative is to launch rapid response optical/IR telescopes in space that would obviate the need
for ground-based facilities. Unfortunately, the cost of even a modest $\frac {1}{2}$-meter aperture telescope
far exceeds installing two dozen much larger instruments located on Earth.

Any instrument dedicated to GRB optical afterglow detection must be robotic with a slew time of tens of seconds in
order to maximize the time overlap with the most variable periods of X-ray and $\gamma$-ray emission. Such a
telescope would be more useful for a broader range of research if the field-of-view (FoV) can be kept large,
at least a square degree. The best example for this argument is the Sloan Digital Sky Survey whose telescope primary has an aperture of 2.5 meters and an FoV of 1.5 square degrees. To complete this picture, the imaging focal
plane could be populated with a 2 $\times$ 2 array of large format silicon CCDs. This would be even more useful
if the instrument could operate as a two-band system with a dichroic splitter to separate R-band and I or
J-band to two different cameras. Such multi-band coverage might better elucidate the origin of 'dark' bursts, whether hidden by optical extinction of dense molecular clouds or redshifted and destroyed by Ly-$\alpha$
absorption edges. Instruments such as described above run counter to the current government
funding trend to shut down many 2-m telescopes in favor of fewer but more powerful 8-m class and larger. Such
policies work well for the majority of astronomical objects which evolve exceedingly slowly with time but are
inappropriate for relatively rare events with durations of minutes or seconds. It also behooves agencies such as NASA
that fund space missions to help organize ground-based programs that will optimize the entire scientific return
on investment.

\section{Acknowledgements}

The authors gratefully acknowledge the extensive help of Fang Yuan, Sarah Yost,
Wiphu Rujopakarn and Eli Rykoff in compiling the lists of optical observations
used for this analysis as well as their comments and suggestions. This research was supported by NSF/AFOSR grant AST-0335588,
NASA grant NNG-04WC41G, NSF grant AST-0407061 and the Michigan Space Grant Consortium.

\clearpage

\begin{deluxetable}{llrcccccccc}
\tablewidth{0pt}
\tablecaption{GRB Afterglow Detections\label{tab:detections}}
\tabletypesize{\scriptsize}
\tablehead{
\colhead{GRB} & \colhead{RA} & \colhead{DEC} & \colhead{Filter} &
\colhead{$A_V$} & \colhead{$A_R$} & \colhead{$\alpha$} & \colhead{$m_{det}$\tablenotemark{a}} & \colhead{$t_{burst}(s)$} &
\colhead{$m_{det}$ @ $t_c$\tablenotemark{b}} & \colhead{Reference} \\
}
\startdata
050318&03:18:51.15&-46:23:43.70&V&0.054&0.043&-0.87&17.80&3230.00&16.445&1\\
050319&10:16:50.76&+43:32:59.90&none&0.036&0.029&-0.88&18.00&1015.00&17.960&2\\
050401&16:31:28.82&+02:11:14.83&none&0.216&0.174&-0.76&18.58&241.35&19.486&3\\
050406&02:17:52.30&-50:11:15.00&V&0.073&0.059&-0.75&19.44&138.00&20.462& 4\\
050416A&12:33:54.60&+21:03:24.00&V&0.098&0.079&*&19.38&115.00&20.516&5\\
050505&09:27:03.20&+30:16:21.50&none&0.071&0.057&*&18.40&1009.00&18.336&6\\
050525&18:32:32.57&+26:20:22.50&none&0.315&0.254&-1.23&16.12&1002.30&15.864&7\\
050607A&20:00:42.79&+09:08:31.50&R&0.516&0.416&-1.00&22.50&960.00&22.115&8\\
050712A&05:10:47.90&+64:54:51.50&V&0.753&0.607&-0.73&17.38&959.00&16.249&9\\
050713A&21:22:09.53&+77:04:29.50&R&1.371&1.106&-0.67&21.41&2963.00&19.478&10\\
050721&16:53:44.53&-28:22:51.80&R&0.894&0.721&-1.29&17.93&1484.00&16.909&11\\
050726&13:20:12.30&-32:03:50.80&V&0.206&0.166&0&17.35&173.00&18.067&12\\
050730&14:08:17.13&-03:46:16.70&R&0.168&0.135&-0.54&17.07&1848.00&16.468&13\\
050801&13:36:35.00&-21:55:41.00&none&0.319&0.257&-1.31&16.93&996.00&16.676&14\\
050802&14:37:05.69&+27:47:12.20&V&0.070&0.057&-0.85&18.35&1463.00&17.581&15\\
050815&19:34:23.15&+09:08:47.47&V&1.457&1.175&*&20.00&117.00&19.764&16\\
050820A&22:29:38.11&+19:33:37.10&Rc&0.146&0.118&-0.97&15.42&1146.00&15.198&17\\
050824&00:48:56.05&+22:36:28.50&none&0.116&0.093&-0.55&18.60&1440.00&18.230&18\\
050908&01:21:50.75&-12:57:17.20&Rc&0.083&0.067&-0.93&18.80&900.00&18.813&19\\
050922C&21:09:33.30&-08:45:27.50&none&0.342&0.276&-1.00&16.00&640.00&16.063&20\\
051109A&22:01:15.31&+40:49:23.31&none&0.630&0.508&-0.65&17.59&1004.00&17.079&21\\
051111&23:12:33.36&+18:22:29.53&none&0.537&0.433&-0.74&16.13&1007.00&15.692&21\\
051117A&15:13:34.09&+30:52:12.70&V&0.080&0.065&-0.35&20.01&210.00&20.706&22\\
051221A&21:54:48.63&+16:53:27.16&R&0.227&0.183&-0.93&20.20&4680.00&18.844&23\\
060108&09:48:01.98&+31:55:08.60&R&0.059&0.047&-0.43&21.84&879.00&21.891&24\\
060110&04:50:57.85&+28:25:55.70&none&2.107&1.699&-0.70&17.90&847.00&16.327&25\\
060111A&18:24:49.00&+37:36:16.10&none&0.094&0.076&*&18.30&173.50&19.555&26\\
060111B&19:05:42.47&+70:22:33.10&none&0.368&0.297&-1.08&18.90&792.00&18.780&27\\
060115&03:36:08.40&+17:20:43.00&Rc&0.441&0.356&0.00&19.10&1190.00&18.612&28\\
060116&05:38:46.28&-05:26:13.14&none&0.873&0.704&-1.09&20.78&926.08&20.134&29\\
060117&21:51:36.13&+59:58:39.10&R&4.292&0.010&-1.70&12.62&502.90&13.132&30\\
060124&05:08:25.50&+69:44:26.00&V&0.449&0.362&0.15&16.79&663.00&16.243&31\\
060203&06:54:03.85&+71:48:38.40&Rc&0.514&0.414&-0.90&19.90&3240.00&18.593&32\\
060204B&14:07:14.80&+27:40:34.00&R&0.059&0.048&-0.80&20.40&3096.00&19.493&33\\
060206&13:31:43.42&+35:03:03.60&r'&0.041&0.033&-1.00&17.80&1036.00&17.740&34\\
060210&03:50:57.37&+27:01:34.40&none&0.309&0.249&-1.30&18.12&835.00&18.008&35\\
060218&03:21:39.68&+16:52:01.82&none&0.471&0.380&*&18.09&858.95&17.826&36\\
060223&03:40:49.56&-17:07:48.36&V&0.385&0.311&-0.75&19.60&935.00&18.856&37\\
060313&04:26:28.40&-10:50:40.10&R&0.230&0.186&-0.13&19.90&1134.00&19.618&38\\
060323&11:37:45.40&+49:59:05.50&none&0.050&0.040&*&18.20&540.00&18.628&39\\
060418&15:45:42.40&-03:38:22.80&Rc&0.743&0.599&-1.20&16.47&2412.00&15.202&40\\
060428B&15:41:25.63&+62:01:30.30&none&0.049&0.040&0.05&19.64&1013.00&19.590&41\\
060502A&16:03:42.48&+66:36:02.50&R&0.109&0.088&-0.45&19.80&2400.00&19.047&42\\
\enddata

\tablenotetext{a}{$m_{det}$ is the measured magnitude at $t_{burst}$ seconds
after the GRB trigger.}
\tablenotetext{b}{$m_{det}$ @ $t_c$ is the inferred value for $m_{det}$ at $t_c$ = 1000 s
after correcting for galactic absorption and average GRB color differences.}

\tablerefs{(1)\citep{2006ApJ...643..276S}
(2)\citep{2006ApJ...640..402Q}
(3)\citep{2005ApJ...631L.121R}
(4)\citep{Schady05_GCN3276}
(5)\citep{2006MNRAS.370.1859D}
(6)\citep{Klotz05_GCN3403}
(7)\citep{2005A&A...439L..35K}
(8)\citep{2006ApJ...645.1315P}
(9)\citep{Poole05_GCN3698}
(10)\citep{2006astro.ph..2387G}
(11)\citep{2006A&A...456..509A}
(12)\citep{McGowan05_GCN3739}
(13)\citep{2006astro.ph..7471P}
(14)\citep{2006ApJ...638L...5R}
(15)\citep{Schady05_GCN3817}
(16)\citep{Holland05_GCN4300}
(17)\citep{2006astro.ph..8183C}
(18)\citep{Lipunov05_GCN3883}
(19)\citep{Torii05_GCN3943}
(20)\citep{Rykoff05_GCN4012}
(21)\citep{2007ApJ...657..925Y}
(21)\citep{2007ApJ...657..925Y}
(22)\citep{2006A&A...456..917R}
(23)\citep{Wren05_GCN4380}
(24)\citep{2006MNRAS.372..327O}
(25)\citep{Li06_GCN4499}
(26)\citep{Klotz06_GCN4483}
(27)\citep{2006A&A...451L..39K}
(28)\citep{Yanagisawa06_GCN4517}
(29)\citep{swan}
(30)\citep{2006A&A...454L.119J}
(31)\citep{Marshall06_GCN4814}
(32)\citep{Bikmaev06_GCN4652}
(33)\citep{Guidorzi06_GCN4661}
(34)\citep{2006ApJ...648.1125M}
(35)\citep{Quimby06_GCN4782}
(36)\citep{Cobb06_GCN4872}
(37)\citep{2007ApJ...654L..21S}
(38)\citep{Zheng06_GCN4930}
(39)\citep{Koppelman06_GCN4977}
(40)\citep{Li06_GCN5027}
(41)\citep{Cenko06_GCN5048}
(42)\citep{2005ApJ...635.1187S}}
\end{deluxetable}

\clearpage

\begin{deluxetable}{llrccccccc}
\tablewidth{0pt}
\tablecaption{GRB Afterglow Non-Detections\label{tab:undetections}}
\tabletypesize{\scriptsize}
\tablehead{
\colhead{GRB} & \colhead{RA} & \colhead{DEC} & \colhead{Filter} &
\colhead{$A_V$} & \colhead{$A_R$}  & \colhead{$m_{lim}$\tablenotemark{a}} & \colhead{$t_{burst}(s)$} &
\colhead{$m_{lim}$ @ $t_c$\tablenotemark{b}} & \colhead{Reference} \\
}
\startdata
050215A&23:13:31.68&+49:19:19.20&none&0.715&0.577&17.40&1080.00&16.765&1\\
050215B&11:37:48.03&+40:47:43.40&V&0.063&0.050&19.50&1797.00&18.582&2\\
050219A&11:05:39.24&-40:40:58.00&V&0.536&0.432&20.70&971.00&19.776&3\\
050219B&05:25:16.31&-57:45:27.31&V&0.109&0.088&19.41&3186.00&18.010&4\\
050223&18:05:32.49&-62:28:21.07&none&0.295&0.238&18.00&2580.00&17.042&5\\
050306&18:49:14.00&-09:09:10.40&none&2.255&1.818&17.50&3600.29&14.708&6\\
050315&20:25:54.10&-42:36:02.20&V&0.159&0.128&18.50&140.19&19.424&7\\
050326&00:27:49.10&-71:22:16.30&V&0.123&0.099&18.91&3313.33&17.467&8\\
050410&05:59:12.90&+79:36:09.20&V&0.369&0.297&19.90&2865.50&18.321&9\\
050412&12:04:25.06&-01:12:03.60&Rc&0.066&0.053&24.90&8336.74&23.235&10\\
050416B&08:55:35.20&+11:10:32.00&r&0.102&0.082&20.00&5160.00&18.671&11\\
050421&20:29:00.94&+73:39:11.40&none&2.693&2.172&18.40&144.29&17.699&12\\
050422&21:37:54.50&+55:46:46.60&V&4.609&3.716&17.90&374.10&13.628&13\\
050502B&09:30:10.10&+16:59:44.30&V&0.098&0.079&21.80&1219.97&21.141&14\\
050509A&20:42:19.70&+54:04:16.20&V&1.981&1.597&18.23&1853.00&15.370&15\\
050509B&12:36:13.67&+28:58:57.00&R&0.064&0.051&21.80&1402.27&21.492&16\\
050528A&23:34:03.60&+45:56:16.80&R&0.533&0.430&20.00&1799.71&19.123&17\\
050713B&20:31:15.50&+60:56:38.40&R&1.548&1.248&21.60&1782.43&19.913&18\\
050714B&11:18:48.00&-15:32:49.90&R&0.181&0.146&20.00&3387.74&18.927&19\\
050716&22:34:20.40&+38:40:56.70&R&0.358&0.289&19.80&228.10&20.634&20\\
050717&14:17:24.90&-50:32:13.20&V&0.786&0.634&18.71&128.00&19.076&21\\
050724&16:24:44.37&-27:32:27.50&V&2.032&1.639&18.84&1663.00&16.011&22\\
050803&23:22:38.00&+05:47:02.30&V&0.246&0.198&18.80&235.01&19.245&23\\
050813&16:07:57.00&+11:14:52.00&V&0.185&0.149&18.15&152.00&18.987&24\\
050814&17:36:45.39&+46:20:21.60&V&0.093&0.075&18.00&217.00&18.658&25\\
050819&23:55:01.20&+24:51:36.50&R&0.406&0.327&21.60&7864.99&19.706&26\\
050820B&09:02:25.03&-72:38:44.00&R&0.417&0.336&15.30&4716.00&13.785&27\\
050822&03:24:26.70&-46:02:01.70&V&0.049&0.040&19.50&138.24&20.545&28\\
050826&05:51:01.58&-02:38:35.80&V&1.944&1.568&19.00&155.00&18.063&29\\
050904&00:54:50.79&+14:05:09.42&V&0.200&0.161&18.90&214.00&19.462&30\\
050906&03:31:11.75&-14:37:18.10&R&0.220&0.177&19.70&470.02&20.097&31\\
050911&00:54:37.70&-38:50:57.70&R&0.034&0.028&21.00&2160.00&20.387&32\\
050915A&05:26:44.80&-28:00:59.27&R&0.086&0.070&21.00&1098.14&20.859&33\\
050915B&14:36:26.50&-67:24:36.50&V&1.292&1.041&21.40&9360.00&17.998&34\\
050922B&00:23:13.20&-05:36:16.40&V&0.122&0.098&20.10&3169.50&18.691&35\\
050925&20:13:54.24&+34:19:55.20&R&7.510&6.056&19.00&197.86&14.175&36\\
051001&23:23:48.80&-31:31:17.00&R&0.051&0.041&21.50&1175.90&21.336&37\\
051006&07:23:13.52&+09:30:24.48&V&0.218&0.176&18.80&207.00&19.369&38\\
051008&13:31:29.30&+42:05:59.00&R&0.039&0.031&22.60&3822.34&21.550&39\\
051016&08:11:16.30&-18:17:49.20&V&0.293&0.236&19.10&114.91&20.041&40\\
051016B&08:48:27.60&+13:39:25.50&Rc&0.123&0.099&15.70&105.41&17.311&41\\
051021B&08:24:11.80&-45:32:30.80&V&3.852&3.106&19.00&178.00&16.050&42\\
051105&17:41:03.28&+34:59:03.60&V&0.112&0.090&20.00&9566.50&17.762&43\\
051109B&23:01:50.21&+38:40:46.00&R&0.557&0.449&21.00&5436.29&19.264&44\\
051117B&05:40:43.00&-19:16:26.50&R&0.185&0.149&20.80&2885.76&19.846&45\\
051221B&20:49:35.10&+53:02:12.20&R&4.543&3.663&18.20&281.66&15.500&46\\
051227&08:20:58.11&+31:55:31.89&Rc&0.140&0.113&17.70&3944.16&16.544&47\\
060105&19:50:00.60&+46:20:58.00&V&0.568&0.458&18.00&191.00&18.280&48\\
060109&18:50:43.50&+31:59:29.70&V&0.478&0.386&19.00&204.00&19.320&49\\
060202&02:23:22.88&+38:23:04.30&R&0.157&0.126&21.50&252.29&22.421&50\\
060211A&03:53:32.80&+21:29:21.00&V&0.637&0.514&19.00&283.00&18.912&51\\
060211B&05:00:17.20&+14:56:58.90&R&1.349&1.088&22.10&2257.63&20.393&52\\
060219&16:07:21.10&+32:18:56.30&V&0.108&0.087&18.60&120.10&19.693&53\\
060223B&16:56:58.80&-30:48:46.00&R&1.301&1.049&13.70&326.59&13.501&54\\
060306&02:44:23.00&-02:08:52.80&V&0.118&0.096&18.40&193.00&19.122&55\\
060312&03:03:06.12&+12:50:03.50&none&0.585&0.472&18.30&1270.94&17.646&56\\
060319&11:45:33.80&+60:00:39.00&R&0.073&0.059&21.00&5238.43&19.682&57\\
060403&18:49:21.80&+08:19:45.30&V&4.251&3.428&19.25&5066.50&13.356&58\\
060413&19:25:07.70&+13:45:27.30&V&6.472&5.219&19.20&1160.00&12.205&59\\
060421&22:54:32.63&+62:43:50.07&V&4.236&3.416&17.70&285.00&14.008&60\\
060427&08:17:04.40&+62:40:18.30&V&0.165&0.133&18.50&333.00&18.761&61\\
060428A&08:14:10.98&-37:10:10.30&V&4.128&3.328&19.10&271.00&15.554&62\\
060501&21:53:29.90&+43:59:53.40&none&0.951&0.767&17.40&426.82&17.280&63\\
060502B&18:35:45.89&+52:37:56.20&none&0.145&0.117&20.00&719.71&20.133&64\\
060507&05:59:51.70&+75:14:56.60&R&0.514&0.414&19.20&3828.38&17.766&65\\
\enddata

\tablenotetext{a}{$m_{lim}$ is the magnitude upper limit at $t_{burst}$ seconds
after the GRB trigger.}
\tablenotetext{b}{$m_{lim}$ @ $t_c$ is the inferred value for $m_{lim}$ at $t_c$ = 1000 s
after correcting for galactic absorption and average GRB color differences.}

\tablerefs{
(1)\citep{Smith05_GCN3021}
(2)\citep{Roming05_GCN3026}
(3)\citep{Schady05_GCN3039}
(4)\citep{Poole05_GCN3050}
(5)\citep{Smith05_GCN3056}
(6)\citep{Klotz05_GCN3084}
(7)\citep{Rosen05_GCN3095}
(8)\citep{Holland05_GCN3150}
(9)\citep{Boyd05_GCN3230}
(10)\citep{Kosugi05_GCN3263}
(11)\citep{Berger05_GCN3283}
(12)\citep{Rykoff05_GCN3304}
(13)\citep{McGowan05_GCN3317}
(14)\citep{Cenko05_GCN3357}
(15)\citep{Poole05_GCN3394}
(16)\citep{Wozniak05_GCN3414}
(17)\citep{Monfardini05_GCN3503}
(18)\citep{lin05_GCN3593}
(19)\citep{Malesani05_GCN3614}
(20)\citep{Guidorzi05_GCN3625}
(21)\citep{Hurkett05_GCN3633}
(22)\citep{Chester05_GCN3670}
(23)\citep{Brown05_GCN3759}
(24)\citep{Retter05_GCN3788}
(25)\citep{Retter05_GCN3799}
(26)\citep{Bikmaev05_GCN3831}
(27)\citep{Jelinek05_GCN3854}
(28)\citep{Page05_GCN3859}
(29)\citep{Mangano05_GCN3884}
(30)\citep{Cucchiara05_GCN3923}
(31)\citep{Fox05_GCN3931}
(32)\citep{Tristram05_GCN3965}
(33)\citep{Cenko05_GCN3981}
(34)\citep{Cobb05_GCN3994}
(35)\citep{Pasquale05_GCN4028}
(36)\citep{Guidorzi05_GCN4035}
(37)\citep{Tristram05_GCN4055}
(38)\citep{Norris05_GCN4061}
(39)\citep{Rumyantsev05_GCN4087}
(40)\citep{Boyd05_GCN4096}
(41)\citep{Torii05_GCN4112}
(42)\citep{Retter05_GCN4126}
(43)\citep{Brown05_GCN4200}
(44)\citep{Huang05_GCN4231}
(45)\citep{Chen05_GCN4285}
(46)\citep{Klotz05_GCN4386}
(47)\citep{Yanagisawa05_GCN4418}
(48)\citep{Ziaeepour06_GCN4429}
(49)\citep{Pasquale06_GCN4455}
(50)\citep{Monfardini06_GCN4630}
(51)\citep{Hurkett06_GCN4736}
(52)\citep{Sharapov06_GCN4925}
(53)\citep{Breeveld06_GCN4798}
(54)\citep{Torii06_GCN4826}
(55)\citep{Angelini06_GCN4848}
(56)\citep{Schaefer06_GCN4860}
(57)\citep{Guziy06_GCN4896}
(58)\citep{Poole06_GCN4951}
(59)\citep{Boyd06_GCN4958}
(60)\citep{Goad06_GCN4985}
(61)\citep{Mangano06_GCN5006}
(62)\citep{Mangano06_GCN5014}
(63)\citep{Rykoff06_GCN5041}
(64)\citep{Zhai06_GCN5057}
(65)\citep{Halpern06_GCN5086}}
\end{deluxetable}

\clearpage

\begin{figure}
\plotone{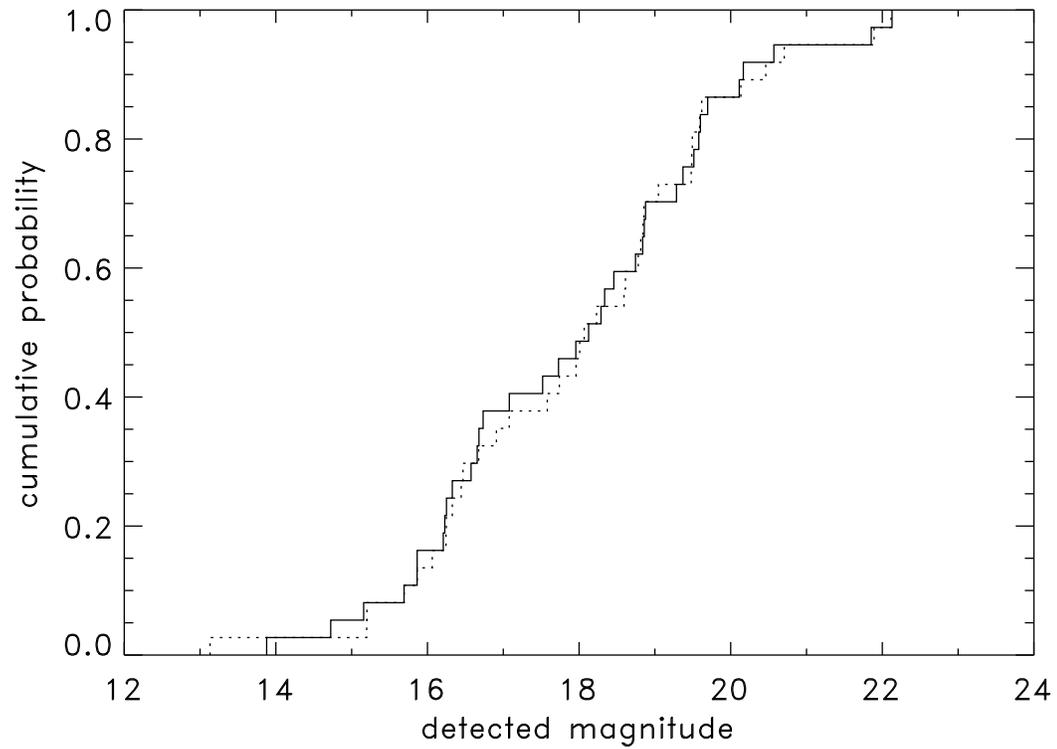}
\caption{The cumulative distributions of afterglow magnitudes for 37 detected GRBs transformed to $t_c = 1000$ s according to a power-law extrapolation. The solid line shows the distribution using a value of $\alpha$ computed individually for each burst; the dotted line represents the similar distribution when $\alpha$ is set to a fixed value of -0.70 for all events. There is no apparent statistical difference between these curves.
 }
\end{figure}

\clearpage

\begin{figure}
\plotone{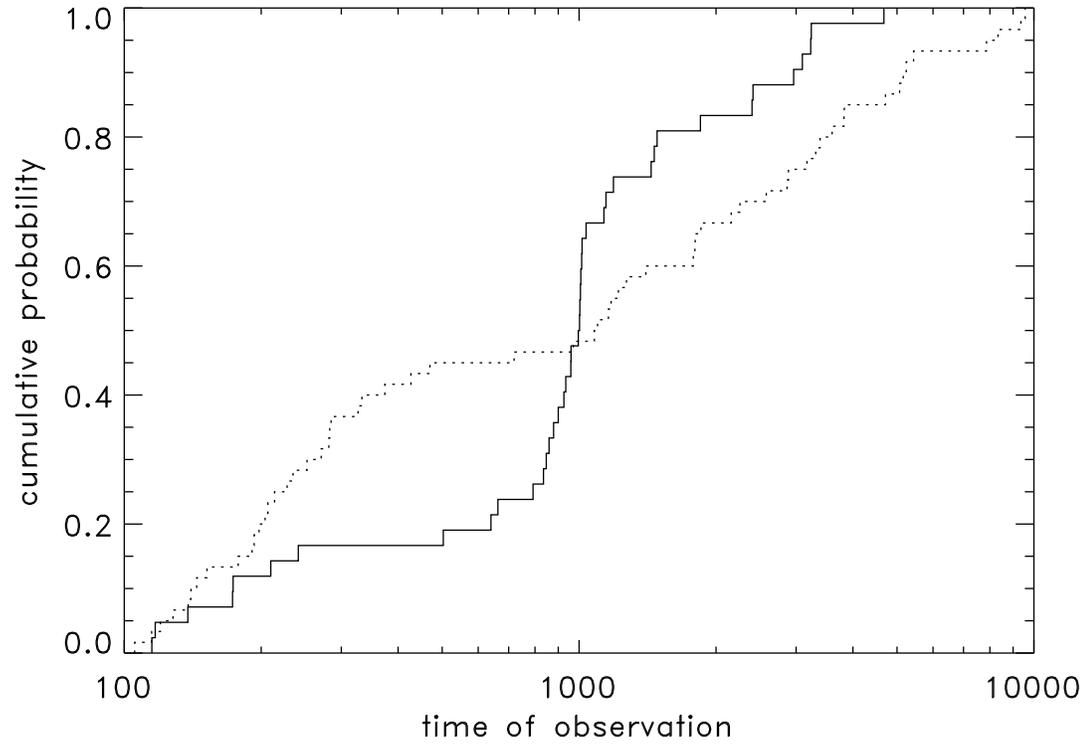}
\caption{The cumulative distributions for the time of observation for detected afterglows (solid line) and undetected afterglows (dotted line) relative to the burst onset. The apparent step function for the detected events at 1000 seconds is an artifact of the selection criteria. Note that the undetected GRBs have a `best' limiting magnitude at a median time also close to 1000 s. This is not a selection effect.
 }
\end{figure}

\clearpage

\begin{figure}
\plotone{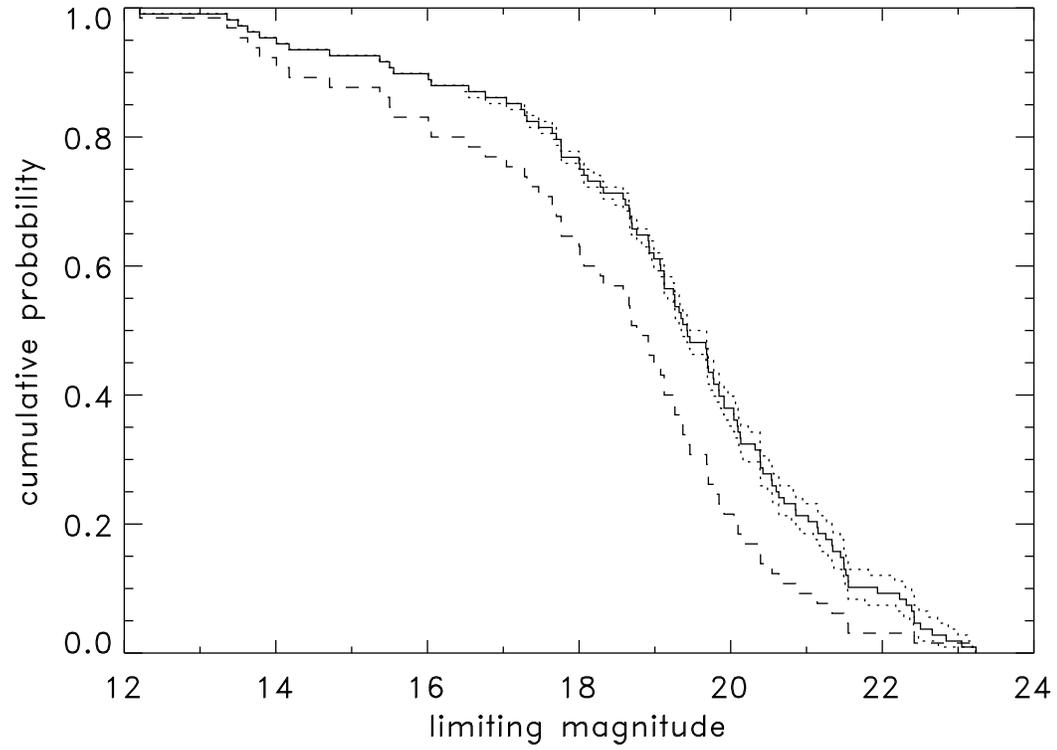}
\caption{The cumulative distribution for observation limiting magnitudes. The
dashed line corresponds to the distribution for the best sensitivity for each non-detection; the solid line is the
estimate obtained for all observations, both detections and non-detections, using the iterative technique described in the text. The dotted lines show the first and third quartile distributions obtained in the Monte Carlo process. These clearly bracket the median quite closely.
}
\end{figure}

\clearpage

\begin{figure}
\plotone{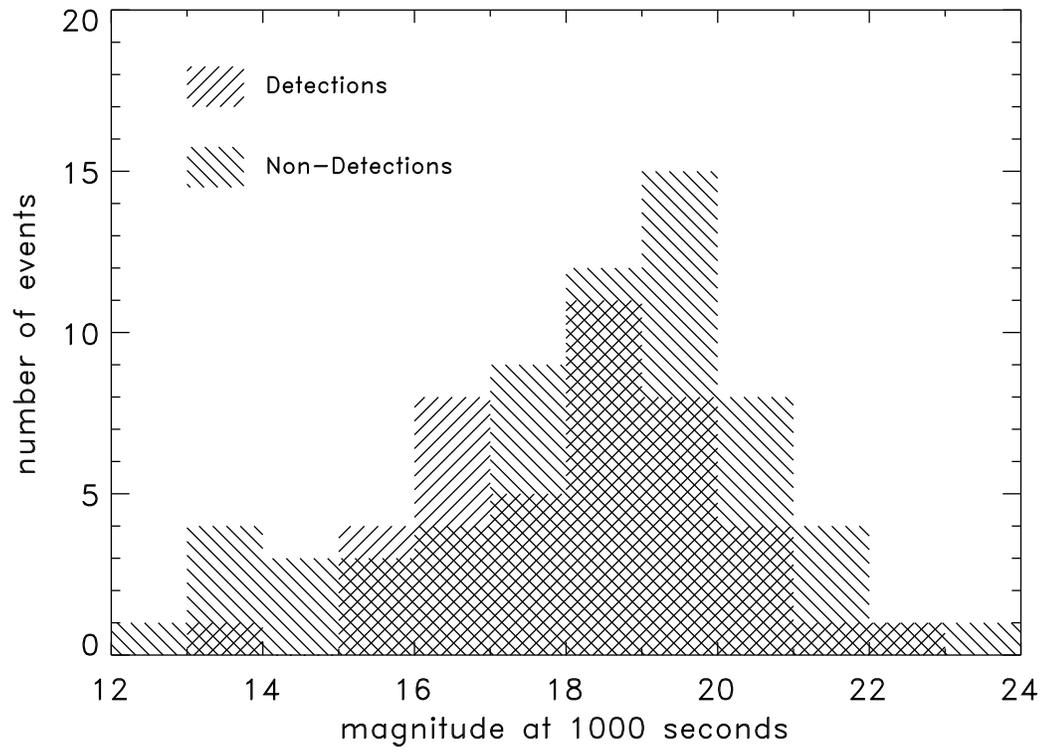}
\caption{Histogram of GRB optical afterglow detections and non-detections transformed to
$t_c$ = 1000 s. Despite a large overlap region, a substantial number of non-detections occur at
limiting magnitudes deeper than most detections.
 }
\end{figure}

\clearpage

\begin{figure}
\plotone{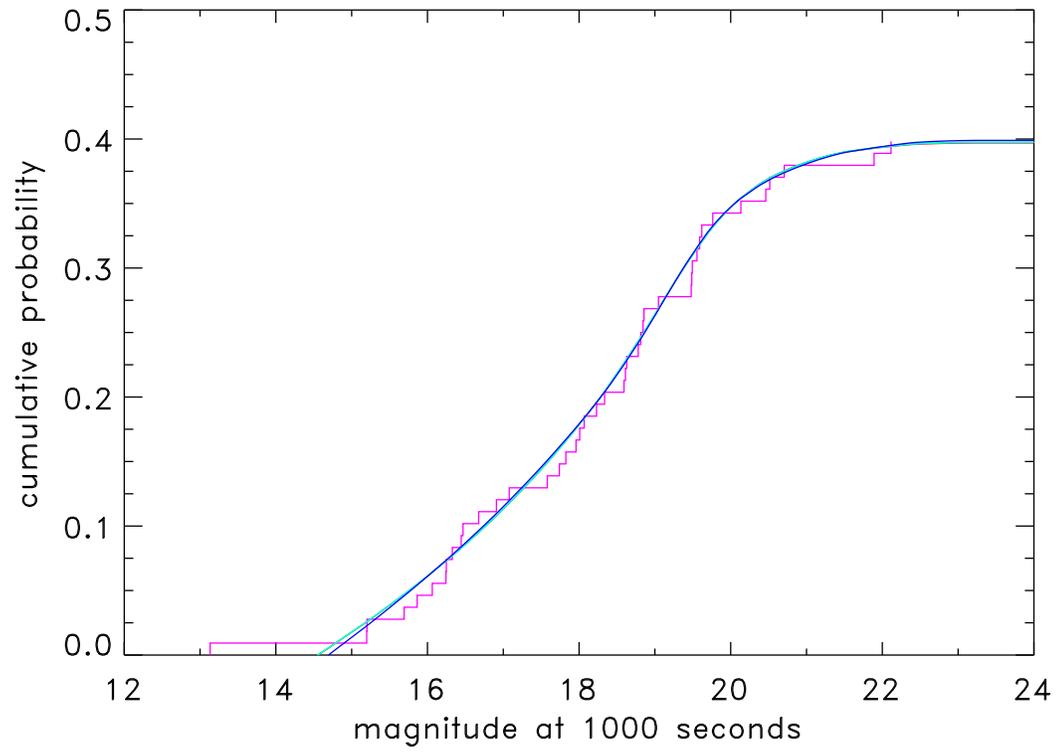}
\caption{The cumulative distribution of detected GRB optical afterglows. The violet
``staircase'' line shows the experimentally observed distribution. The least square estimates are shown in red,
green, cyan and blue, corresponding respectively to 4, 5, 6 and 7 degrees of freedom of the b-spline representation.
 }
\end{figure}

\clearpage

\begin{figure}
\plotone{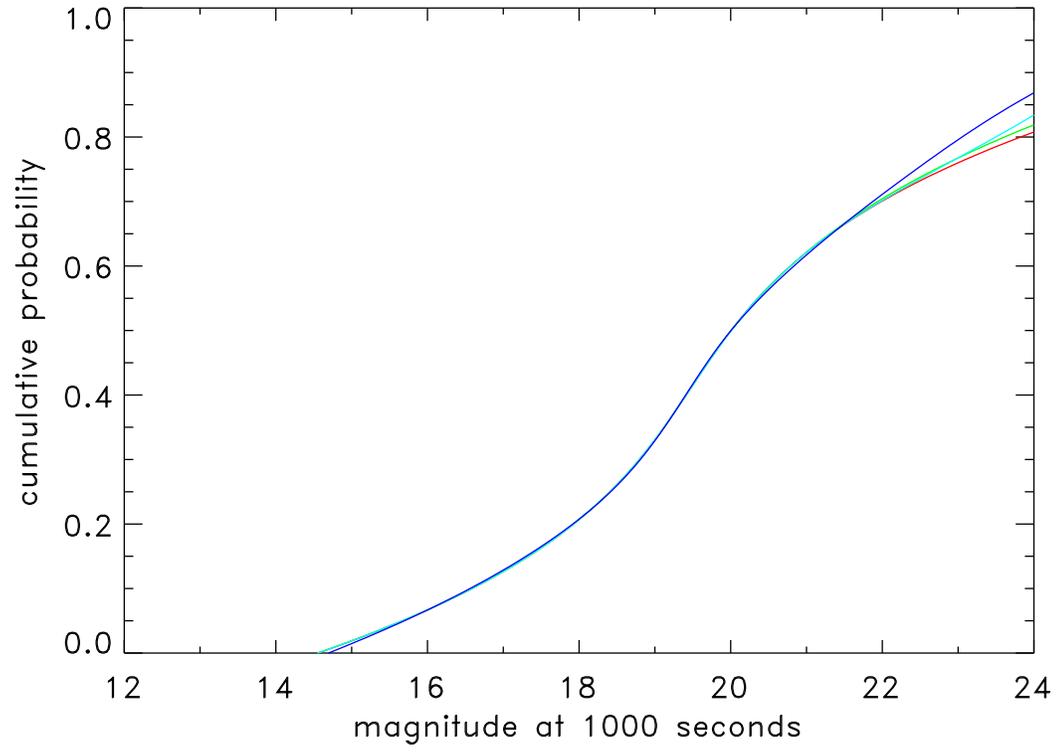}
\caption{The intrinsic cumulative GRB apparent optical afterglow distribution. The colors,
red, green, cyan and blue, correspond respectively to b-spline curves with
4, 5, 6 and 7 degrees of freedom. Crudely speaking, 71\% of all GRB afterglows
have $m_R < 22.1$ at $t_{obs} = 1000~s$, the dimmest GRB optically detected.
 }
\end{figure}

\clearpage

\begin{figure}
\plotone{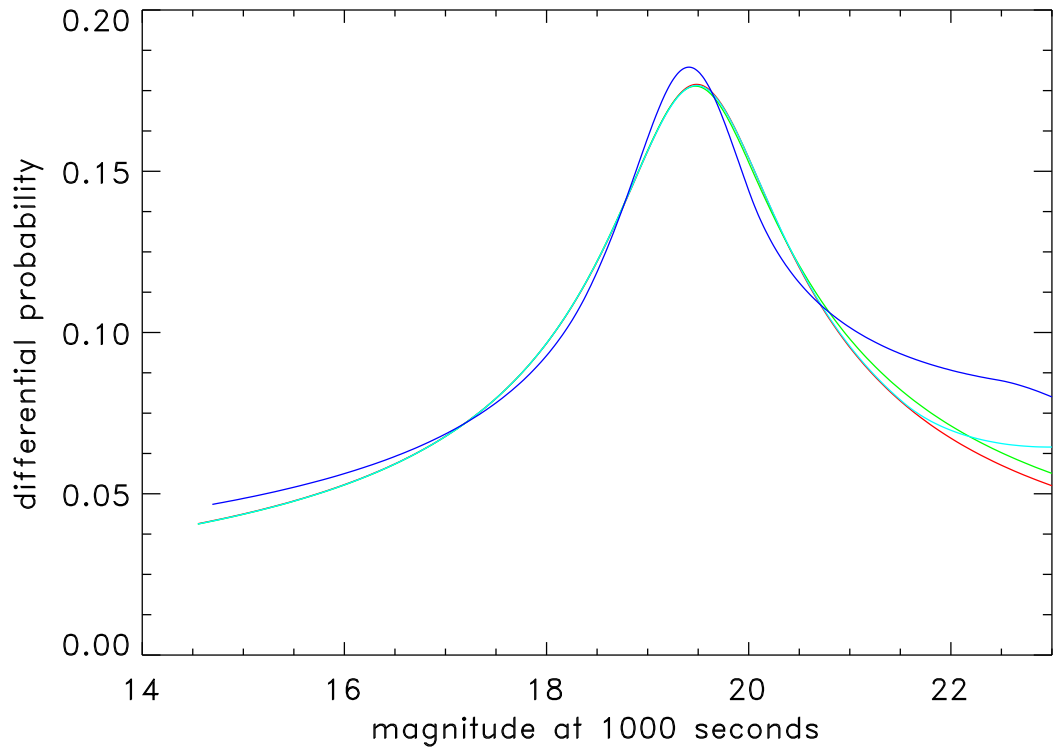}
\caption{The differential GRB apparent optical afterglow distribution. The colors,
red, green, cyan and blue, correspond respectively to b-spline curves with
4, 5, 6 and 7 degrees of freedom. The peak at $m_{det} \approx 19.5$ seems to be an unavoidable feature.
 }
\end{figure}

\end{document}